%% file: Matching-and-Fair-Division.tex
\relax
\documentclass[letterpaper]{article} % DO NOT CHANGE THIS
\usepackage{fullpage}
\usepackage{times}  % DO NOT CHANGE THIS
\usepackage{helvet} % DO NOT CHANGE THIS
\usepackage{courier}  % DO NOT CHANGE THIS
\usepackage[hyphens]{url}  % DO NOT CHANGE THIS
\usepackage{graphicx} % DO NOT CHANGE THIS
\usepackage[numbers]{natbib}  % DO NOT CHANGE THIS AND DO NOT ADD ANY OPTIONS TO IT
\usepackage{caption} % DO NOT CHANGE THIS AND DO NOT ADD ANY OPTIONS TO IT
\usepackage{amsmath,amssymb,amsthm,amsfonts,nicefrac}
\usepackage{cleveref}
\usepackage{algorithm}
\usepackage[noend]{algpseudocode}

\newcommand{\commentsymbol}{//}
\algrenewcommand\algorithmiccomment[1]{\hfill \commentsymbol{} #1}
\makeatletter
\newcommand{\LineComment}[2][\algorithmicindent]{\Statex \hspace{#1}\commentsymbol{} #2}
\makeatother

\DeclareMathOperator*{\argmin}{argmin}
\DeclareMathOperator*{\argmax}{argmax}
\newtheorem{lemma}{Lemma}
\newtheorem{theorem}{Theorem}
\theoremstyle{definition}
\newtheorem{definition}{Definition}

\newtheorem{example}{Example}
\newtheorem{proposition}{Proposition}
\newcommand{\ceil}[1]{\left\lceil{#1}\right\rceil}
\newcommand{\floor}[1]{\left\lfloor{#1}\right\rfloor}
\newcommand{\set}[1]{\left\{#1\right\}}
\renewcommand{\hat}{\widehat}

\newcommand{\bbR}{\mathbb{R}}

\newcommand{\bbZ}{\mathbb{Z}}

\newcommand{\Nr}{N^r}
\newcommand{\nr}{n^r}
\newcommand{\Nl}{N^\ell}
\newcommand{\nl}{n^\ell}
\newcommand{\dl}{d^\ell}
\newcommand{\dr}{d^r}

\newcommand{\Ml}{M^\ell}
\newcommand{\Mr}{M^r}

\newcommand{\Ar}{A^r}

\newcommand{\ul}{u^\ell}
\newcommand{\ur}{u^r}
\newcommand{\rl}{\sigma^\ell}
\newcommand{\rr}{\sigma^r}
\newcommand{\mmsl}{MMS^{\ell}}
\newcommand{\mmsr}{MMS^{r}}
\newcommand{\Sl}{S^{\ell}}
\newcommand{\Sr}{S^{r}}
\newcommand{\Bl}{B^{\ell}}
\newcommand{\Br}{B^{r}}

\usepackage[dvipsnames]{xcolor}

\newcount\Comments
\usepackage{color}
\definecolor{darkgreen}{rgb}{0,0.5,0}
\definecolor{purple}{rgb}{1,0,1}
\newcommand{\kibitz}[2]{\ifnum\Comments=1\textcolor{#1}{#2}\fi}

\usepackage[switch]{lineno}
\title{Two-Sided Matching Meets Fair Division\thanks{A preliminary version appeared in the proceedings of the 30th International Joint Conference on Artificial Intelligence (IJCAI-21).}}
\author{
Rupert Freeman\\University of Virginia\\\texttt{FreemanR@darden.virginia.edu}
\and
Evi Micha\\University of Toronto\\\texttt{emicha@cs.toronto.edu}
\and
Nisarg Shah\\University of Toronto\\\texttt{nisarg@cs.toronto.edu}
}
\date{}

\begin{document}
	
\maketitle

\begin{abstract}
  We introduce a new model for two-sided matching which allows us to borrow popular fairness notions from the fair division literature such as envy-freeness up to one good and maximin share guarantee. In our model, each agent is matched to multiple agents on the other side over whom she has additive preferences. We demand fairness for each side separately, giving rise to notions such as double envy-freeness up to one match (DEF1) and double maximin share guarantee (DMMS). We show that (a slight strengthening of) DEF1 cannot always be achieved, but in the special case where both sides have identical preferences, the round-robin algorithm with a carefully designed agent ordering achieves it. In contrast, DMMS cannot be achieved even when both sides have identical preferences.
\end{abstract}

\section{Introduction}\label{sec:intro}
\input{1-intro}

\section{Preliminaries}\label{sec:prelims}
\input{2-preliminaries}

\section{Double Envy-Freeness Up To One Match}\label{sec:envy-free}
\input{3-envy-free}

\section{Double Maximin Share Guarantee} 
\input{4-DMMS}

\section{Discussion}
\input{6-discussion}

\section*{Acknowledgements}
Nisarg Shah acknowledges support from an NSERC Disocvery Grant.

\bibliographystyle{named}
\bibliography{abbshort,ultimate}

\cleardoublepage
\onecolumn
%\leftlinenumbers
\appendix
\section*{Appendix}

\section{Additional Material From \Cref{sec:envy-free}}\label{app:proof-of-theor-d=2}
\input{app-1-arb_pref_d=2}

\section{SD-DEF1 and DMMS for $d=2$}\label{app:theor:def1-and-dmms}
\input{app-4-SD-DEF1-MMS-d=2}

\section{Other Versions of DMMS}\label{sec:other-dmms}
\paragraph{DMMS without degree constraints}
When we defined the maximin share of agent $i$, we only considered matchings  that respect the degree constraint. One may ask how much this constraint affects the maximin share, relative to an alternative definition where we maximize over all matchings in which each $i \in \Nl$ is matched with each $j \in \Nr$ at most once, but degree constraints are not required to hold. 

Relative to this stronger version of MMS, a MMS matching over $\Nl$ that respects the degree constraint may only give a $\Omega(1/n)$ approximation.
Consider an example with $n$ even and $d=2$, and $\ul(j)=n$ for any $j \in [n/2-2]$, while $\ul(j)=1$ for any remaining agents on the right. This means that $n-2$ agents on the left can be matched with one of the first agents on the right, and gain utility equal to $n$, while $2$ agents can be matched with all the agents in $\{n/2-1,...,n-1\}$, and gain utility equal to $n/2+1$. But as $d=2$, this means that under a matching that respects the degree constraint, there is an agent that has utility equal to $2$. Hence, the approximation is  $\Omega(1/n)$.

\paragraph{A weaker version of DMMS}
We have seen that DMMS can not be achieved, even when agents have identical valuations, $n^\ell=n^r=n$, and $d^\ell=d^r=d$. However, consider the following relaxation of the MMS share (for simplicity, consider the MMS share of an agent $i \in N^\ell$). Instead of maximizing over all matchings that satisfy the degree constraint, suppose that we instead maximize over all partitions $\pi = \{ \pi_1, \ldots, \pi_k \}$ of $N^r$ such that none of the resulting bundles contain more than $d^\ell$ agents from $N^r$. That is, denoting by $\Pi$ the space of all partitions of $N^r$ that respect the degree constraint, the \emph{weak MMS share} of agent $i$ is defined as
	\begin{align*}
	wMMS_i^\ell = \max\limits_{\pi \in \Pi} \min\limits_{\pi_j \in \pi} \ul_i(\pi_j).
	\end{align*}
When all agents in $N^\ell$ and $N^r$ have identical valuations and $n$, $d$ are common, weak MMS can always be achieved. First, find the partitions $\pi^\ell$ and $\pi^r$ that maximize the minimum value of any bundle, subject to respecting the degree constraint and breaking ties between partitions by selecting those that minimize the total number of bundles. 
Therefore, $u_i(\pi^r_j) \ge wMMS_i^\ell$ for all $i \in N^\ell$ and $\pi^r_j \in \pi^r$, and $u_j(\pi_i) \ge wMMS_j^r$ for all $j \in N^r$ and $\pi^\ell_i \in \pi^\ell$. Note that since each bundle formed by $\pi^\ell$ and $\pi^r$ can have at most $d$ members, there must by exactly $\lceil \frac{n}{d} \rceil$ bundles in each of $\pi^\ell$ and $\pi^r$ (if there were fewer bundles, one must contain more than $d$ agents, and if there were more bundles then it would be possible to weakly increase the value of the lowest-valued bundle by merging two bundles). 
Now we can match each agent in $\pi_i^\ell$ to each agent in $\pi_i^r$ for all $i \in \{ 1, \ldots, \lceil \frac{n}{d} \rceil \}$. 
This matching respects degree constraints and satisfies weak MMS, by the definition of $\pi^\ell$ and $\pi^r$.

\section{DMMS \& DEF1}\label{sec:dmms-def1}

In this section we explore the relationship between DMMS and DEF1. 

\begin{proposition}
The existence of a DEF1 matching does not necessarily imply the existence of a DMMS matching.
\end{proposition}
\begin{proof}
We revisit the instance of the proof of Theorem \ref{theor:DMMS-not-guaranteed} and set: 
\begin{itemize}
		\item $M^\ell_0=M^\ell_5=\{0, 4, 5\}$
		\item $M^\ell_2=\{0, 3, 6\}$
		\item $M^\ell_1=M^\ell_6=\{1,2,6\}$
		\item $M^\ell_3=\{1, 3, 5\}$
		\item $M^\ell_4=\{2, 3, 4\}$
	\end{itemize}
It is easy to verify that the matching is DEF1, and the proposition follows.  
\end{proof}

\begin{proposition}
A DMMS matching may not be DFE1.
\end{proposition}
\begin{proof}
Consider the instance that $n=10$, $d=3$, and $\ul(0)=\ul(1)=\ul(2)=2$, and $\ul(j)=1$ for any $j \in \{3,...,9\}$, while $\ur(i)=1$ for any $i \in [n]$. Hence, $\mmsl=3$ and $\mmsr=3$, and  notice that any complete matching is DMMS. Now, consider the matching that $\Ml_0=\Ml_1=\Ml_2=\{0,1,2\}$. Clearly, $M$ is not DEF1. 
\end{proof}

In the next theorem we prove that, given the existence of a DMMS matching that minimizes the number of agents that receive utility equal to $\mmsl$ in $\Nl$ and simultaneously minimizes the number of agents that receive utility equal to $\mmsr$ in $\Nr$, we are guaranteed the existence of a DEF1 matching.

\begin{theorem}\label{theor:DMMS-implies-DEF1}
	A DMMS matching that minimizes the number of agents that receive utility equal to $\mmsl$ in $\Nl$ and simultaneously minimizes the number of agents that receive utility equal to $\mmsr$ in $\Nr$ is also DEF1.
\end{theorem}
\begin{proof}
	We show  that when a  complete matching is MMS over $\Nl$ and minimizes the agents  on the left that receive utility equal to $\mmsl$, it should also be EF1 over it. 

  For contradiction, suppose that there is a  complete matching which is MMS over $\Nl$, minimizes the agents  on the left that receive utility equal to $\mmsl$,  but it is not EF1 over the left side. Let $\Ml_{i'}=\argmin_{\Ml_i \in \Ml} \ul(\Ml_i)$ and $i''$ be an agent such that $i'$ envies $i''$ for more than one matches. Notice that $\ul(\Ml_{i'})= \mmsl$.  As $\ul(\Ml_{i'}) < \ul(\Ml_{i''})$, there is at least one pair $j \in \Ml_{i'}$ and $j' \in \Ml_{i''}$ such that $\ul(j) < \ul(j')$. Now consider a different matching $\hat{M}^{\ell}$ such that $\hat{M}^{\ell}_{i}=\Ml_{i}$, for every $i\neq \{i',i''\}$, while  $\hat{M}^{\ell}_{i'}=\Ml_{i'} \setminus j \cup j'$, and $\hat{M}^{\ell}_{i''}=\Ml_{i''} \setminus j' \cup j$. Then, $\ul(\hat{M}^{\ell}_{i'}) > \ul (\Ml_{i'} )$, as $\ul(j) < \ul(j')$. In addition, we have
	\begin{align*}
	\ul(\hat{M}^{\ell}_{i''}) \geq \ul(\hat{M}^{\ell}_{i''} \setminus j)  > \ul(\Ml_{i'}) =\mmsl
	\end{align*}
	where the third inequality follows from the fact that the matching is not EF1, as $\hat{M}^{\ell}_{i''} \setminus j=\Ml_{i''}\setminus i'$. Thus, we conclude in a matching in which the number of agents that receive utility $\mmsl$ has been increased which is a contradiction. 
  
 The theorem follows by making similar arguments for $\Nr$.
\end{proof}

One may wonder whether it is always possible, given a DMMS matching, to find a DMMS matching that satisfies the condition of Theorem~\ref{theor:DMMS-implies-DEF1}. If this were true, then the existence of a DMMS matching would imply the existence of a DEF1 matching. However, we now show that simultaneously minimizing the number of agents that receive utility equal to $\mmsl$ in $\Nl$ and the number of agents that receive utility equal to $\mmsr$ in $\Nr$ may not be possible. 

\begin{proposition}
It does not always exist a DMMS matching that minimizes the agents that receive utility equal to $\mmsl$ in $\Nl$ and concurrently minimizes the agents that receive utility equal to $\mmsr$ in $\Nr$.
\end{proposition}
\begin{proof}
Consider the instance that $n=7$, $d=3$,  $\ul(0)=\ul(1)=3$,  $\ul(2)=\ul(3)=\ul(4)=1$, and  $\ul(5)=\ul(6)=0$, while $\ur(0)=9$, $\ur(1)=\ur(2)=\ur(3)=\ur(4)=3$, and $\ur(5)=\ur(6)=0$. Notice that $\mmsr=9$, and it should exist three agents $j$, $j'$, and $j''$ such that $\Mr_j=\Mr_{j'}=\Mr_{j''}=\{0,5,6\}$, and hence $\Ml_0=\Ml_5=\Ml_6=\{j,j',j''\}$. Moreover, notice that $\mmsl=3$, and the number of agent that receive utility equal to $3$ is minimized if all  agents have two matches among the first $5$ agents, except from one that is matched with $2$, $3$ and $4$. Hence, we see that there is no way three agents on the left to be matched with the same three agents on the right, and the proposition follows.  However, notice that the following matching:
\begin{itemize}
\item $\Mr_0= \{0 ,5 ,6 \}$
\item $\Mr_1= \{1 ,2 ,3 \}$
\item $\Mr_2= \{4 ,1 ,2 \}$
\item $\Mr_3= \{4 ,1 ,3 \}$
\item $\Mr_4= \{4 ,2 ,3 \}$
\item $\Mr_5= \{0 ,5 ,6 \}$
\item $\Mr_6= \{0 ,5 ,6 \}$
\end{itemize}
is DMMS and DFE1.
\end{proof}

\section{Simulations}\label{app:simulations}
\input{5-simulations}
\end{document}

%% file: 1-intro.tex
Consider a group of agents seeking to divide some number of indivisible goods amongst themselves. Each agent has a utility function describing the value that they have for every possible bundle of goods, and each agent may have a different utility function. This is a canonical resource allocation problem that arises in estate division, partnership dissolution, and charitable donations, to name just a few. A central goal is to find an allocation of the goods that is \emph{fair}. 

One desirable notion of fairness is \emph{envy-freeness}~\cite{Fol67}, which requires that no agent prefer another agent's allocation of goods to her own. This is a compelling definition but, due to the discrete nature of the problem, cannot always be satisfied. Instead, we must consider relaxed versions, with one popular relaxation being \emph{envy-freeness up to one good} (EF1)~\cite{LMMS04,Bud11}, which requires that any pairwise envy can be eliminated by removing a single good from the envied agent's allocation. An allocation satisfying EF1 always exists for a broad class of agent utility functions~\cite{LMMS04}. 

While quite general, the resource allocation model fails to capture some allocation settings that we might be interested in. In particular, it does not allow for the possibility of \emph{two-sided} preferences, in which agents have preferences over ``goods,'' but also ``goods'' have preferences over agents. For instance, when college courses are allocated to students, it is reasonable to assume that students have preferences over the courses they take, and that teachers in charge of courses also have preferences over the students they accept (perhaps measured by prerequisites or GPA).\footnote{Assigning students to courses has been studied before~\citep{Bud11,OSB10,BC12}, but these papers typically only consider the preferences of the students.} As another example, consider the problem of matching social services to vulnerable individuals\footnote{\url{http://csse.utoronto.ca/social-needs-marketplace}}, where individuals have preferences over the services they receive, and service providers have preferences over the individuals they serve (perhaps based on demographics, location, or synergy with existing clients).

Allowing for two-sided preferences is immediately reminiscent of the matching literature. In two sided matching, it is generally assumed that each agent has ordinal preferences over the other side, and a matching is sought that is in some sense \emph{stable} to individual or group deviations. 
It is well-known that stability is closely related to envy-freeness in the sense that a one-to-one matching is stable if and only if it eliminates \emph{justified envy}~\citep{AS03}, a requirement specifying that any envy that $i$ may feel for $j$'s match is ``justified'' by $j$'s match preferring $j$ to $i$. In many-to-one settings, the notions remain tightly connected, with a stable matching being one that eliminates justified envy and has no waste.  Justified envy-freeness has also been studied as in its own right in the many-to-one setting \citep{wu2018lattice,yokoi2020envy}.

Justified envy, and therefore stability, fundamentally rely on the idea that the less an agent is preferred by the other side, the lower her own entitlement should be. However, in some applications, it may be desirable to provide \emph{equal} entitlements or moral claims to agents regardless of how valued they are by the agents on the other side. For example, instructors may prefer students with high GPA over students with low GPA, but it is not clear that universities should adopt such a policy in their course scheduling (and, in fact, usually do not).
Therefore, while stability is a valuable notion in many settings, in this work we consider two-sided preferences while incorporating traditional notions from fair division.

\paragraph{Our contributions.} We introduce and study a two-sided resource allocation setting in which we have two groups of agents, where each agent has preferences over agents on the other side. Each agent must be ``matched'' to a subset of agents on the other side, subject to a maximum degree constraint. 
Our goal is to find a many-to-many matching that provides fairness to both sets of agents simultaneously. 
The standard resource allocation setting is a special case of our model in which each good can be matched to at most one agent, each agent can be matched to any number of goods, and goods are indifferent to which agent they are assigned to.

As a natural tradeoff between expressiveness and succinctness, we restrict our attention to additive preferences, in which the utility for being matched to a group of agents is equal to the sum of utilities for being matched to each agent in the group individually. While conceptually simple, additive preferences have led to a rich body of work in fair division. We focus primarily on the case in which all agents on the same side have the same degree constraint, and the total maximum degree on both sides is equal. In this case, it is reasonable to seek a \emph{complete} matching, which saturates the degree constraints of all the agents on both sides. 

We begin by considering \emph{double envy-freeness up to one match} (DEF1), requiring that EF1 hold for both sets of agents simultaneously. 
We show that a complete matching satisfying (a slight strengthening of) DEF1 does not always exists, but in the special case where both sides have identical ordinal preferences, it exists and can be computed efficiently using a carefully designed round robin algorithm.

We also ask whether it is possible to find matchings that satisfy \emph{double maximin share guarantee} (DMMS), a two-sided version of the maximin share guarantee. Even when both sides have identical preferences, a complete DMMS matching may not exist, in contrast to the one-sided 
setting in which an MMS allocation is guaranteed to exist when agents have identical preferences. In general, we show that approximate DMMS and approximate DEF1 are incompatible, although in the special case where the degree constraint is equal to two we can achieve exact versions of both simultaneously.

\paragraph{Related work.} Most related to our work is that of \citet{patro2020fairrec}, who draw on the resource allocation literature to guarantee fairness for both producers and consumers on a two-sided platform. However, in their model, producers are indifferent between the customers; thus, only one side has interesting preferences. Other work~\cite{chakraborty2017fair,suhr2019two} has focused on guaranteeing fairness in two-sided platforms over time, rather than in a one-shot setting. Of particular note is the work of \citet{gollapudi2020almost}, who consider two-sided EF1 in a dynamic setting, but obtain positive results primarily for symmetric binary valuations, a much more restrictive class of valuations than we consider. \citet{tadenuma2011partnership} studies envy minimization in two-sided matching subject to other notions, including stability, but focuses on ordinal notions of envy and restricts attention to one-to-one matchings.

The theories of matching and fair division each have a rich history. Traditional work in matching theory has focused on one-to-one or many-to-one matchings, beginning with the seminal work of~\citet{GS62} and finding applications in areas such as school choice~\cite{AS03,APRS05,HKN11}, kidney exchange~\cite{SRSU+06}, and the famous US resident-to-hospital match.\footnote{\url{https://www.nrmp.org/}} We note that EF1 as a condition becomes vacuous whenever a set of agents has a maximum degree constraint of one, so we focus instead on the more general case of many-to-many matchings. This case has also been well-explored in the matching literature~\cite{Roth84,Soto99,RS92,echenique2004theory}, although that literature focuses on stability notions, which have a very different flavor to our guarantees.

Our work draws extensively on notions from the fair division literature, particularly envy-freeness and its relaxations~\cite{Fol67,Bud11,LMMS04} and the maximin share guarantee~\cite{Bud11}. Prior work has studied the satisfiability of these properties in the resource allocation setting~\cite{CKMP+19,PW14,KPW16}, including the house allocation setting in which each agent is ``matched'' to a single item~\citep{aigner2019envy,beynier2019local,gan2019envy}, but, to our knowledge, no work has considered satisfying them on both sides of a market simultaneously.

%% file: 2-preliminaries.tex
For $n \in \mathbb{N}$, define $[n]=\set{0,\ldots,n-1}$. There are two disjoint groups of agents, denoted $\Nl$ (``left'') and $\Nr$ (``right''), of sizes $\nl$ and $\nr$, respectively. For simplicity of notation, we write $\Nl = [\nl]$ and $\Nr = [\nr]$; when referring to an agent by only its index, the group she belongs to will be clear from context. We use indices $i \in [\nl]$ and $j \in [\nr]$ to refer to agents on the left and right, respectively. We are given degree constraints $\dl_i$ and $\dr_i$ such that each $i \in \Nl$ and each $j \in \Nr$ can be matched to at most $\dl_i$ and $\dr_j$ agents on the opposite side, respectively.   When $\dl_i=\dl_{i'}$ for any $i,i' \in \Nl$ (resp. $\dr_j=d^r_{j'}$ for any $j,j' \in N^r$), we denote by $\dl$ (resp. $\dr$) the common degree constraint of  all agents in $\Nl$ (resp.  $\Nr$).

A (many-to-many) \emph{matching} $M$ is represented as a binary $\nl \times \nr$ matrix, where $M(i,j)=1$ if $i \in \Nl$ and $j\in \Nr$ are matched, and  $M(i,j)=0$ otherwise. With slight abuse of notation, we denote $\Ml_i = \set{j \in \Nr : M(i,j) = 1}$ and $\Mr_j = \set{i \in \Nl : M(i,j) = 1}$ as the sets of agents on the opposite side that agents $i \in \Nl$ and $j \in \Nr$ are matched to, respectively. We say that $M$ is \emph{valid} if it respects the degree constraints, i.e., if $|\Ml_i|\leq \dl_i $ for each $i\in N^l$ and $|\Mr_j| \leq \dr_j $ for each $j\in \Nr$. Hereinafter, we omit the term valid, but will always refer to valid matchings. 
We say that $M$ is \emph{complete} if $\sum_{i \in \Nl} |\Ml_i| = \sum_{j \in \Nr} |\Mr_j| = \min(\sum_{i \in \Nl} \dl_i, \sum_{j \in \Nr} \dr_j)$. That is, a complete matching is one in which either every agent on the left has their degree constraint met exactly, or every agent on the right does.

Each agent $i \in \Nl$ has a valuation function $\ul_i : \Nr \to \bbR_{\ge 0}$ and each agent $j \in \Nr$ has a valuation function $\ur_j : \Nl \to \bbR_{\ge 0}$. When agents $i \in \Nl$ and $j \in \Nr$ are matched, they simultaneously receive utilities $\ul_i(j)$ and $\ur_j(i)$, respectively. We assume that utilities are additive. Thus, with slight abuse of notation, the utilities to agents $i \in \Nl$ and $j \in \Nr$ under matching $M$ are $\ul_i(\Ml_i) = \sum_{j \in \Ml_i} \ul_i(j)$ and $\ur_j(\Mr_j)= \sum_{i \in \Mr_j} \ur_j(i)$, respectively. 

Our main constructive results take only the agents' preference orders as input. For agent $i \in \Nl$ (resp. $j \in \Nr$), we denote by $\rl_i$ (resp. $\rr_j$) a linear order over $\Nr$ (resp. $\Nl$) which is consistent with the valuation function $\ul_i$ (resp. $\ur_j$), i.e., $j \succ_{\rl_i} j'$ whenever $\ul_i(j) > \ul_i(j')$ (resp. $i \succ_{\rr_j} i'$ whenever $\ur_j(i) > \ur_j(i')$).\footnote{Ties among agents with equal utility are broken arbitrarily.} With a slight abuse of notation, we denote with $\rl_i(p)$ (resp. $\rr_j(p)$) the position of alternative $p$ in $\rl_i$ (resp. $\rr_j$).

Inspired by envy-freeness up to one good (EF1) from classical fair division~\cite{Bud11,LMMS04}, we define the following fairness guarantee in our setting. 

\begin{definition}[Double Envy-Freeness Up To $c$ Matches (DEF$c$)]
We say that matching $M$ is envy-free up to $c$ matches (EF$c$) over $\Nl$ if for each pair of agents $i,i' \in \Nl$, there exists $\Sl \subseteq \Ml_{i'}$ with $|\Sl| \le c$ such that $\ul_i(\Ml_i) \ge \ul_i(\Ml_{i'} \setminus \Sl)$. Similarly, we say that it is EF$c$ over $\Nr$ if, for each pair of agents $j,j' \in \Nr$, there exists $\Sr \subseteq \Mr_{j'}$ with $|\Sr| \le c$ such that $\ul_j(\Mr_j) \ge \ul_j(\Mr_{j'}\setminus\Sr)$. We say that $M$ is DEF$c$ if it is EF$c$ over both $\Nl$ and $\Nr$. 
\end{definition}

When an algorithm takes as input only the preference rankings, it must ensure that the matching it returns is DEF$c$ for all possible valuation functions which could have induced the rankings. It is easy to observe that this is equivalent to satisfying the following stronger guarantee which uses the stochastic dominance (SD) relation.
This is akin to the SD-EF1 strengthening of EF1~\citep{FSV20,Aziz20}.
	
\begin{definition}[SD Double Envy-Freeness Up To $c$ Matches (SD-DEF$c$)]
We say that matching $M$ is SD-envy-free up to $c$ matches (SD-EF$c$) over $\Nl$ if, for every $t \in [\nr]$,
\begin{align*}
\textstyle\sum_{p=0}^t M(i,\rl_i(p)) \geq \sum_{p=0}^t M(i',\rl_i(p))-c, \forall i,i' \in \Nl,
\end{align*}
and is SD-EF$c$ over $\Nr$ if, for every $t \in [\nl]$,
\begin{align*}
\textstyle\sum_{p=0}^t M(\rr_j(p),j) \geq \sum_{p=0}^t M(\rr_j(p),j')-c, \forall j,j' \in \Nr.
\end{align*}
$M$ is called SD-DEF$c$ if it is SD-EF$c$ over both $\Nl$ and $\Nr$. 
\end{definition}

Finally, we extend a different fairness notion from classical fair division called the maximin share guarantee (MMS).

\begin{definition}[$\alpha$-Double Maximin Share Guarantee ($\alpha$-DMMS)]
	Let $\mathcal{M}$ denote the set of valid matchings. The \emph{maximin share value} of agent $i \in \Nl$ is defined as   
	\[
	\textstyle\mmsl_i = \max_{M \in \mathcal{M}}\ \min_{i' \in \Nl}\ \ul_i(\Ml_{i'}),
	\]
	and the maximin share value of agent $j \in \Nr$ is defined as
	\[
	\textstyle\mmsr_j = \max_{M \in \mathcal{M}}\ \min_{j' \in \Nr}\  \ur_j(\Mr_{j'}).
	\]
	Given $\alpha \in [0,1]$, matching $M$ is called $\alpha$-maximin share fair ($\alpha$-MMS) over $\Nl$ if $\ul_i(\Ml_i) \geq \alpha \cdot \mmsl _i$ for every $i \in \Nl$, and $\alpha$-MMS over $\Nr$ if $\ur_j(\Mr_j)\geq \alpha \cdot \mmsr_j$ for every $j \in \Nr$. It is called $\alpha$-DMMS if it is $\alpha$-MMS for both $\Nr$ and $\Nr$. When $\alpha=1$, we write DMMS instead of $1$-DMMS.
\end{definition}

The notions of (SD-)DEF1 and DMMS are incomparable to the traditional notions of stability and justified envy-freeness, as the following example shows.

\begin{example}
Suppose that $n^\ell=n^r=4$, i.e., $N^\ell=N^r=\{0,1,2,3\}$, the common degree requirement is $2$, and each side has identical ordinal preference $0 \succ 1\succ 2 \succ 3$ over the other side. The only matching that is stable and eliminates justified envy is the one that matches each $i \in \{0,1\}$ on the left with every $j \in \{0,1\}$ on the right, and each $i \in \{2,3\}$ on the left with every $j \in \{2,3\}$ on the right. Indeed,  if some $i \in \{0,1\}$ on the left is not matched to some $j \in \{0,1\}$ on the right,  then $j$ must be matched to some $i' \in \{2,3\}$ on the left, which would make $(i,j)$ a blocking pair, and $i$ would (justifiably) envy $i'$ for her match with $j$. However, this matching violates DEF1 when, for example, agent 2 on the left has more value for agent 1 on the right than for agents 2 and 3 on the right combined (as this would leave her envious of agents 0 and 1 on the left, even after ignoring their match to agent 0 on the right). Note that this matching also violates DMMS, since each agent on the left could partition those on the right into bundles $\{0,3 \}, \{ 0,3 \}, \{ 1,2 \}, \{ 1,2 \}$, guaranteeing themselves a better bundle than the $\{ 2, 3 \}$ that agents 2 and 3 receive.

On the other hand, any one-to-one matching satisfies \mbox{(SD-)DEF1} and DMMS, but many one-to-one matchings are not stable or free of justified envy.
\end{example}

%% file: 3-envy-free.tex
In this section, we focus on double envy-freeness up to one match, more specifically, its strengthening SD-DEF1. 
We begin in \Cref{sec:impossibility} by presenting an impossibility result that holds even under quite restrictive conditions. Then, in \Cref{sec:possibility}, we present an algorithm that efficiently computes an SD-DEF1 matching whenever both groups of agents have identical ordinal preferences. In \Cref{app:proof-of-theor-d=2}, we present an additional positive result for the case that all agents have maximum degree constraint equal to two, and one side has identical preferences.

\subsection{SD-DEF1 Matchings May Not Exist}
\label{sec:impossibility}

Our first main result says that a complete SD-DEF1 matching may not exist. Observe that without the completeness condition an empty matching is trivially SD-DEF1. The proof 
uses a counterexample in which both sides have the same number of agents (that is, $\Nl=\Nr=n$), all agents have the same degree constraint ($\dl=\dr=d$), and one group of agents have identical preferences ($\ul_i=\ul_{i'}$ for all $i, i' \in \Nl$). Thus, \Cref{thm:imposs-d-ge-3} holds even in this restricted case, and continues to hold for more general settings.

\begin{theorem}\label{thm:imposs-d-ge-3}
	A complete SD-DEF1 matching is not guaranteed to exist.
\end{theorem}
\begin{proof}
	Let $d \ge 3$ and $n=4d$. Let all agents in $\Nl$ have identical preferences over agents in $\Nr$ given by $0 \succ \ldots \succ n$. To define the preferences of agents in $\Nr$ over $\Nl$, let us partition the agents in $\Nr$ into $d$ blocks: $\Br_m=\set{4m,\ldots,4(m+1)-1}$ for $m \in [d]$. We define a preference ranking $\rho_m$ for each block $\Br_m$, and let all agents in the block have this preference ranking. The first three rankings $\rho_0$, $\rho_1$, and $\rho_2$ are shown below. The agents not shown in these rankings (marked ``remaining agents'') can appear in an arbitrary order at the end. Rankings $\rho_3,\ldots,\rho_{d-1}$ can be completely arbitrary. 
	\begin{itemize}
		\item $\rho_0 = 0 \succ 1 \succ 2 \succ 3 \succ \ldots $
		\item $\rho_1 = 0 \succ 1 \succ 4 \succ 5 \succ\ldots$
		\item $\rho_2 = 2 \succ 3 \succ 4 \succ 5 \succ\ldots $
	\end{itemize}

	Once again, we claim that this instance does not admit any complete SD-DEF1 matching. Suppose for contradiction that it does. 
	
	Under such a matching, each agent in $\Nl$ must be matched to exactly one agent in $\Br_m$ for every $m \in [d]$. Suppose for contradiction that this is not true; let $m$ be the smallest index for which the statement fails. Then, because the total degree in $\Br_m$ is $4d = |\Nl|$, there must exist agents $i,i' \in \Nl$ such that $i$ is matched to no agent from $\Br_m$, $i'$ is matched to at least two agents from $\Br_m$, and both $i$ and $i'$ are matched to exactly one agent from $\Br_{m'}$ for each $m' < m$. This would violate SD-EF1 with respect to agents $i$ and $i'$. 

	Notice that each agent in $\Br_0$ must be matched to exactly one agent from $S_0 = \set{0, 1, 2, 3} \subset \Nl$. If this is not true, then because each agent in $S_0$ is matched to exactly one agent from $\Br_0$ and $|S_0| = |\Br_0| = 4$, we must have agents $j,j' \in \Br_0$ such that agent $j$ is matched to at least two agents from $S_0$ while agent $j'$ is matched to none of them. This would violate SD-EF1 with respect to agents $j$ and $j'$. By a similar reasoning, every agent in $\Br_1$ must be matched to exactly one agent in $S_1 = \set{0, 1, 4, 5} \subset \Nl$, and every agent in $\Br_2$ must be matched to exactly one agent in $S_2 = \set{2,3,4,5} \subset \Nl$.
	
	Consider agent $j \in \Br_0$ that agent $4$ from $\Nl$ is matched to. By the first observation above, there must be a unique agent $i \in S_0$ who is also matched to agent $j$. We take two cases. 
	
	\medskip\noindent\textbf{Case 1:} Suppose $i \in \set{0,1}$. Then, agent $j$ is matched to two agents from $\set{0,1,4} \subset S_1$ --- agents $i$ and $4$. In contrast, the agent $j' \in \Br_1$ who is matched to agent $5$ from $S_1$ is not matched to any agent from $\set{0,1,4} \subset S_1$ by the second claim above. Hence, SD-EF1 is violated with respect to agents $j$ and $j'$, which is a contradiction.
	
	\medskip\noindent\textbf{Case 2:} Suppose $i \in \set{2,3}$. Then, agent $j$ is matched to two agents from $\set{2,3,4} \subset S_2$ --- agents $i$ and $4$. In contrast, the agent $j' \in \Br_2$ who is matched to agent $5$ from $S_2$ is not matched to any agent from $\set{2,3,4} \subset S_2$ by the third claim above. Hence, SD-EF1 is violated with respect to agents $j$ and $j'$, which is a contradiction.
\end{proof}

A natural question, that we leave open, is whether the impossibility continues to hold when we relax SD-DEF1 to DEF1. We have found no counterexamples (even for SD-DEF1) via simulation; the counterexample for SD-DEF1 is carefully crafted but relies on the strength of SD-DEF1.

\subsection{Identical Ordinal Preferences on Both Sides}
\label{sec:possibility}

\Cref{thm:imposs-d-ge-3} says that a complete SD-DEF1 matching is not guaranteed to exist even under quite restrictive conditions. It is natural to ask whether there exist any settings in which a complete SD-DEF1 matching can be guaranteed. In \Cref{app:proof-of-theor-d=2}, we establish the existence of a complete SD-DEF1 matching by restricting the degree bound (with still only one side having identical preferences). In this section, we consider a different restriction: when agents on both sides have identical preferences, i.e., $\rl_i = \rl_{i'}$ for all $i,i' \in \Nl$ and $\rr_j = \rr_{j'}$ for all $j,j' \in \Nr$. 

\begin{theorem}\label{thm:general}
When $\nl \dl = \nr \dr$ and both groups of agents have identical ordinal preferences, a complete SD-DEF1 matching always exists and can be computed efficiently. 
\end{theorem}

The proof of \Cref{thm:general} follows from a series of lemmas. In the main text we focus on the simple case for which $\Nl=\Nr=n$ and 
$\dl=\dr=d$.
\Cref{thm:general} follows from progressively reducing the general case to this simple case.

We denote by $\rl$ and $\rr$ the ordinal preferences of the agents in $\Nl$ and $\Nr$, respectively. Without loss of generality, assume that $\rl = \rr = 0 \succ \ldots \succ n-1$. We want to find an SD-DEF1 matching under which each agent is matched to exactly $d$ agents on the opposite side. A natural idea is to let agents on one side pick agents on the other side in a round-robin fashion. That is, we construct an ordering $R$ over agents on one side, and these agents take turns according to $R$ in a cyclic fashion with each agent, in her turn, making one match to her most preferred agent (i.e. lowest indexed agent) on the opposite side who has less than $d$ matches so far. A standard argument from classical fair division shows that regardless of the ordering $R$, the resulting matching will be SD-EF1 over over the side that does the picking.\footnote{As observed by \citet{biswas2018fair}, the standard round robin algorithm is not EF1 when agents have cardinality constraints, but EF1 is retained provided that agents have identical preferences.} However, as the example below shows, not all orderings $R$ lead to a matching that also satisfies SD-EF1 over the other side. 

\begin{example}
	\label{ex:rr-violates-sd-def1}
	Consider the case where $n=5$ and $d=2$. Suppose the ordering $R$ has agents on the left choose in the order 0, 1, 2, 3, 4. Then, agent $0$ on the right will be matched to agents $0$ and $1$ on the left, while agent $1$ on the right will be matched to agents $2$ and $3$ on the left. 
	SD-EF1 is violated as agent $1$ significantly envies agent $0$ on the right side.
\end{example}

We now show that when $R$ is carefully designed, SD-EF1 can also be satisfied over the other side, resulting in SD-DEF1. \Cref{alg:order} takes as input parameters $a \in [n]$ and $x \in \set{d,n-d}$, and for any choices of these parameters, constructs an ordering $R$ over the agents on (say) the left side. \Cref{alg:coprime} then uses this ordering to run the round-robin procedure while respecting the degree constraints. Example~\ref{ex:alg-1-and-2} demonstrates these algorithms.

\begin{algorithm}[htb!]
	\caption{Round-Robin-Ordering$(n,a,x)$ }\label{alg:order}
	\begin{algorithmic}[1]
		\LineComment[0\dimexpr\algorithmicindent]{$x$ and $n$ coprime, so $x^{-1} \pmod n$ exists}
		\For {$i \in [n]$}
			\State $R(p) = a + p x^{-1} \pmod n$
		\EndFor
		\State \textbf{return} $R$
	\end{algorithmic}
\end{algorithm}
\begin{algorithm}[htb!]
	\caption{{Restricted-Round-Robin-Coprime}($n, d$)}\label{alg:coprime}  
	\begin{algorithmic}[1]
		\State Choose $a \in \set{0,\ldots,n-1}$ and $x \in \set{d,n-d}$ 
		\State $R=$Round-Robin-Ordering$(n,a,x)$
		\LineComment[0\dimexpr\algorithmicindent]{Round-robin with ordering $R$ over agents on the left}	
		\State $M(i,j)=0, \forall i,j \in [n]$
		\For {$j \in [n], t \in [d]$}		
			\State $M(R(j \cdot d + t \pmod n),j)=1$ \label{matching}
		\EndFor
		\State \textbf{return} $M$
	\end{algorithmic}
\end{algorithm}

\begin{example}
	\label{ex:alg-1-and-2}
	Consider the same instance as Example~\ref{ex:rr-violates-sd-def1}, with $n=5$ and $d=2$. Suppose we choose $a=3$ and $x=d=2$. Then the round robin ordering returned by Algorithm~\ref{alg:order} is $R(0)=3+0=3$ (setting $i=0$), $R(1)=3+3=1$ ($i=3$), $R(2)=3+1=4$ ($i=1$), $R(3)=3+4=2$ ($i=4$), and $R(4)=3+2=0$ ($i=2$), with all addition performed mod $n$. That is, agents on the left choose in order $3, 1, 4, 2, 0$. This results in the matching $M_3^\ell=\{ 0, 2 \}$, $M_1^\ell=\{ 0, 3 \}$, $M_4^\ell=\{ 1, 3 \}$, $M_2^\ell=\{ 1, 4 \}$, and $M_0^\ell=\{ 2, 4 \}$ (equivalent to the formula provided directly in Line~\ref{matching} of Algorithm~\ref{alg:coprime}). The fact that this is SD-EF1 over $\Nl$ is easy to check. Examining the matching, note that $M_0^r=\{ 1, 3 \}=M_4^\ell$, $M_1^r=\{ 2, 4 \}=M_0^\ell$, $M_2^r=\{ 0, 3 \}=M_4^\ell$, $M_3^r=\{ 1, 4 \}=M_2^\ell$, and $M_4^r=\{ 0, 2 \}=M_3^\ell$. That is, the matchings received by agents on the right are the same as those received by agents on the left, up to a cyclic shift. For this matching, SD-EF1 over $\Nl$ immediately implies SD-EF1 over $\Nr$.
\end{example}

The next result shows that for any choices of the parameters, the resulting matching is SD-DEF1. The idea of the proof is to show that the structure in Example~\ref{ex:alg-1-and-2} holds in general: for any allowed choice of $(a,x)$, the set of bundles received by agents on the right is the same as the set of bundles received by agents on the left, thus inheriting SD-EF1 from the fact that the matching is constructed by agents on the left choosing in round robin sequence. 

\begin{lemma}\label{theor:coprime}
	When $\nl=\nr=n$ and $\dl=\dr=d$ are coprime and both groups of agents have identical ordinal preferences, \Cref{alg:coprime} efficiently computes a complete SD-DEF1 matching.
\end{lemma}
\begin{proof}
To avoid the $\pmod n$ notation in this proof, we will treat integers as belonging to the ring $\bbZ/n\bbZ$ of integers modulo $n$. Thus, addition, multiplication, and multiplicative inverses will be modulo $n$.
Note that $x^{-1}$ exists because $x \in \set{d,n-d} = \set{d,-d}$ is coprime with $n$.

We claim that the ordering $R$ constructed in \Cref{alg:order} is a valid ordering over the agents in $\Nl$. 
Notice that because $x \in \set{d,-d}$ is coprime with $n$, $(p \cdot x^{-1})_{i \in [n]} = [n]$. Thus, each position in the ordering $R$ is mapped to exactly one agent. Because agents on the left take $d$ turns in a cyclic fashion, it is convenient to think of an \emph{extended} ordering $R$ which is the original $R$ concatenated with itself $d$ times: one can check that this still obeys $R(p) = a + p x^{-1}$ for all $p \in [n d]$. 

Next, we argue that the matching returned is a valid complete matching. Notice that during the round-robin, $d$ agents on the left that are consecutive in the ordering pick a given agent on the right before moving on to the next lowest indexed agent on the right. Further, each agent on the left gets $d$ turns. Hence, it is easy to see that every agent is matched to exactly $d$ agents on the opposite side. 

As mentioned earlier, the fact that the returned matching $M$ is SD-EF1 over $\Nl$ follows directly from the standard round-robin argument in classical fair division: given any pair of agents $i,i' \in \Nl$, if we ignored the first turn taken by $i'$, then in each round agent $i$ would get a turn before agent $i'$ does, and hence, would not envy agent $i'$ in the SD sense. It remains to show that $M$ is also SD-EF1 over $\Nr$. We show that for each agent $j \in \Nr$, there exists an agent $i \in \Nl$ such that $\Mr_j = \Ml_i$. SD-EF1 over $\Nr$ will then follow from SD-EF1 over $\Nl$ given that $\rl=\rr$.

Let us focus on agent $j \in \Nr$. Because agents on the right are picked from lowest-indexed to highest-indexed, agent $j$ is picked by the $d$ agents from $\Nl$ who appear consecutively in the (extended) ordering $R$ at indices $jd+t$ for $t \in [d]$. Given that $R(p) = a+px^{-1}$ for all $p \in [nd]$, we immediately have 
$\Mr_j=\set{a+(jd+t)x^{-1} : t=[d]}$.

Next, let us focus on agent $i \in \Nl$. If she is matched to some agent $j \in \Nr$ in a particular turn, then from the observation above, it must be that $i = a+(jd+t)x^{-1}$ for some $t \in [d]$. Solving this for $j$, we get that $j= ((i-a)x-t)d^{-1}$. Varying $t \in [d]$ in this equation gets us the $d$ agents on the right that agent $i$ is matched to:
$\Ml_i=\set{((i-a)x-t)d^{-1} : t=[d]}$.

To show that for each $j \in \Nr$, there exists $i \in \Nl$ with $\Ml_i = \Mr_j$, we take two cases. 	
	
If $x = n-d = -d$, then $x^{-1} = (-d)^{-1} = -d^{-1}$. In this case, it is easy to check that taking $i=j$ suffices as 
$\Mr_j =\Ml_j =\set{a-j-td^{-1} : t = [d]}$.

If $x=d$, then $\Mr_j= \set{j+a+td^{-1} : t=[d]}$,	
while
\begin{align*}
\Ml_i &=\set{i-a-td^{-1} : t=[d]} \\
&= \set{i-a-(d-1-t)d^{-1} : t=[d]}.
\end{align*}	
Notice that $\Mr_j$ coincides with $\Ml_{j+2a-d^{-1}+1}$.
\end{proof}

\Cref{alg:coprime} executes round-robin with the left side taking turns, and allows freely choosing $a \in [n]$ and $x \in \set{d,n-d}$ to decide their ordering. Note that if the right side takes turns instead, the algorithm still produces a complete SD-DEF1 matching. However, this extension does not find any new matchings. When $x=n-d$, the matching produced is symmetric ($\Ml_i = \Mr_i$ for all $i \in [n]$), and thus the same regardless of which side takes turns. When $x=d$, the allocations on one side are cyclic shift of the allocations on the other side. Hence, any matching produced by the right side taking turns can also be produced by the left side taking turns with appropriately chosen $(a,x)$. 

What about allowing choices of $x$ other than just $d$ and $n-d$? At least for $n=7$, $d=3$, and $a=0$, it is easy to check by hand that no other choices of $x$ produce an SD-DEF1 matching.  On the other hand, could it be that some of the $2n$ choices of $(a,x)$ are redundant and lead to the same matching as other choices? The following result shows that in every instance, all $2n$ choices lead to different matchings. 

\begin{proposition}\label{prop:diff-matchings}
	For any inputs $n$ and $d$ to \Cref{alg:coprime}, the $2n$ possible choices of $(a,x)$ result in distinct matchings.
\end{proposition}
\begin{proof}
	Using the same reasoning as in the proof of \Cref{theor:coprime}, for each $p \in [n]$, we have 
	\[
	R(p)=a+p\cdot x^{-1} = \begin{cases}
	a + p \cdot d^{-1}, &\mbox{ if $x=d$},\\
	a - p \cdot d^{-1} = a+n-p \cdot d^{-1}, &\mbox{ if $x=n-d$}.
	\end{cases}
	\]
	
	Note that if $(a,x)$ generates an ordering $(a,a_1,\ldots,a_{n-1})$, then $(a,\hat{x})$ with $\hat{x} \neq x$ generates the ordering $(a,a_{n-1},\ldots,a_1)$. Moreover, using different choices of $a$ while fixing the choice of $x$ generates orderings that are cyclically shifted versions of one another. 
	
	Let us denote with $R$ and $\hat{R}$ the orderings returned by \Cref{alg:order} using  $(a,x)$ and $(\hat{a}, \hat{x})$, respectively, and let $M$ and $\hat{M}$ denote the matchings returned by \Cref{alg:coprime} under orderings $R$ and $\hat{R}$, respectively. Suppose for contradiction that $(a,x) \neq (\hat{a},\hat{x})$ but $M = \hat{M}$. 
	
	Let $R=(a, a_1,...,a_{d-1},...,a_{n-1})$. Then, because agents in $\Nl$ pick agents on the opposite side in a round-robin fashion, we immediately have that $\Mr_0=\set{a,a_1,...,a_{d-1}}$. If $d < n/2$, then $a_{n-1} \notin \Mr_1$, while if $d > n/2$, then $a \in \Mr_1$. Note that $d=n/2$ is not possible because $d$ and $n$ are coprime. We now consider two cases.
	
	\medskip\noindent\textbf{Case I: $x=\hat{x}$.} As $R$ and $\hat{R}$ are just cyclically shifted, it is easy to observe that $\Mr_0 = \hat{M}^r_0$ implies $a=\hat{a}$, which is a contradiction.
	
	\medskip\noindent\textbf{Case II: $x\neq \hat{x}$.} Let $R'$ be the ordering obtained with $(a,\hat{x})$. Then, $R'=(a, a_{n-1},...,a_{d-1},...,a_1)$. Further, note that $\hat{R}$ is a cyclically shifted version of $R'$, and we want $\hat{M}^r_0 = \Mr_0 = \set{a,a_1,\ldots,a_{d-1}}$. It is easy to notice that the only way to obtain this is by setting $\hat{a} = a_{d-1}$, which will induce $\hat{R}=(a_{d-1}, ...,a_1, a, a_{n-1},...,a_{d})$, making $a_{d-1},\ldots,a_1,a$ the first $d$ agents in the order. If $d < n/2$, then we have $a_{n-1} \in \hat{M}^r_1$ but $a_{n-1} \notin \Mr_1$, while if $d > n/2$, then we have $a \notin \hat{M}^r_1$ but $a \in \Mr_1$. In either case, we have that $\Mr_1 \neq \hat{M}^r_1$, which is a contradiction. 
\end{proof}

Given \Cref{prop:diff-matchings}, one may be tempted to conjecture that these $2n$ choices generate all complete SD-DEF1 matchings. However, as the following example shows, this is not the case, leaving open the question of characterizing the set of all complete SD-DEF1 matchings.
\begin{example}\label{example:non-captured-matching}
	Consider the instance with $n=5$, $d=2$, and identical ordinal preferences on both sides. Let us focus on the following complete matching $M$.
	\begin{itemize}
		\item $\Ml_0=\Mr_0=\{0,2\}$
		\item $\Ml_1=\Mr_1=\{1,3\}$
		\item $\Ml_2=\Mr_2=\{0,4\}$
		\item $\Ml_3=\Mr_3=\{1,4\}$
		\item $\Ml_4=\Mr_4=\{2,3\}$
	\end{itemize}
	
	It is easy to verify that this matching is SD-DEF1.  Without loss of generality, we can assume that agents in $\Nl$ pick agents on the opposite side in a  round robin fashion. Then, agents $2$ and $3$ should be in the last two positions of the round-robin ordering as they are matched to the least preferred agent in $\Nr$. 
	
	\begin{table}[h]
	\centering
	\begin{tabular}{c c c}
		& $x=d$ & $x=n-d$  \\
		$a=0$ & $[0, 3, 1, 4, 2]$ & $[0, 2, 4, 1, 3]$\\
		$a=1$ & $[1, 4, 2, 0, 3]$ & $[1, 3, 0, 2, 4]$\\
		$a=2$ & $[2, 0, 3, 1, 4]$ & $[2, 4, 1, 3, 0]$\\
		$a=3$ & $[3, 1, 4, 2, 0]$ & $[3, 0, 2, 4, 1]$\\
		$a=4$ & $[4, 2, 0, 3, 1]$ & $[4, 1, 3, 0, 2]$
	\end{tabular}
	\caption{Round-robin orderings returned by \Cref{alg:order} for different choices of $(a,x)$ when $n=5$ and $d=2$.}
	\label{table:rr-orderings}
	\end{table}
	
	\Cref{table:rr-orderings} shows the round robin orderings produced by all the choices of $(a,x)$. The reader can verify that neither places agents $2$ and $3$ in the last two positions. Hence, $M$ is not returned by any of the choices. 
\end{example}

The proof of \Cref{thm:general} continues by reducing the case where $n$ and $d$ are not coprime to the coprime case. Letting $g = \gcd(n,d)$, we divide both sides into $g$ sub-groups of $n' = n/g$ agents each. Then, we run \Cref{alg:coprime} a total of $g^2$ times to match agents from each sub-group on the left to $d' = d/g$ agents from each sub-group on the right. This matches each agent with exactly $d$ agents from the opposite side. Note that we allow each of the $g^2$ calls to \Cref{alg:coprime} to use arbitrary choices of $a$ and $x$. Nonetheless, we show that the resulting complete matching must be SD-DEF1. 

\begin{lemma}
	\label{lem:not-coprime}
	When $\nl=\nr=n$, $\dl=\dr=d$, and both groups of agents have identical ordinal preferences, a complete SD-DEF1 matching always exists and can be computed efficiently.
\end{lemma}
\begin{proof}
	Consider \Cref{alg:no-coprime}.
	
	\begin{algorithm}[htb!]
		\caption{{Restricted-Round-Robin}($n,d$) }\label{alg:no-coprime}
		\begin{algorithmic}[1]
			\State $g=\gcd(n,d)$, $n'=n/g$, and $d'=d/g$
			\For {$k,m \in [g]$}
				\State $\hat{i}=\set{n'\cdot k + i : i \in [n']}$	
				\State $\hat{j}=\set{n'\cdot m + j : j \in [n']}$
				\State  $M(\hat{i},\hat{j})$={Restricted-Round-Robin-Coprime}($n',d'$)
			\EndFor
			\State \textbf{return} $M$
		\end{algorithmic}
	\end{algorithm}	
		
	For $k,m \in [g]$, define the sets $\Bl_k = \set{n' \cdot k + i : i \in [n']}$ and $\Br_m = \set{n' \cdot m + j : j \in [n']}$ as used in \Cref{alg:no-coprime}. Note that \Cref{alg:no-coprime} calls \Cref{alg:coprime} on each pair $(\Bl_k,\Br_m)$, and as a result, each agent in $\Bl_k$ is matched to exactly $d'$ agents in $\Br_m$, and vice-versa. 
		
	We want to show that the overall matching $M$ produced by \Cref{alg:no-coprime} is SD-DEF1. Let us first show that it is SD-EF1 over $\Nl$. Consider two arbitrary agents $i,i' \in \Nl$. Given that their preference rankings are $0 \succ \ldots \succ n-1$, to show that SD-EF1 holds for these two agents, we need to show that for all $t \in [n]$,
	\begin{align*}
	\sum_{p=0}^t M(i,p) \geq \sum_{p=0}^t M(i',p)-1.
	\end{align*} 
	
	Fix $t \in [n]$. Because $\set{\Br_m : m \in [g]}$ forms a partition of $\Nr = [n]$, there exists a unique $m \in [g]$ such that $t \in \Br_m$. For each $m' < m$, each of agents $i$ and $i'$ are marched to exactly $d'$ agents in $\Br_{m'}$. Hence, $\sum_{p=0}^{n'\cdot m-1} M(i,p) = \sum_{p=0}^{n'\cdot m-1} M(i',p) = m \cdot d'$. Thus, it remains to show that
	\begin{align}\label{eqn:desired}
	\sum_{p \in \Br_m : p \le t} M(i,p) \geq \sum_{p \in \Br_m : p \le t} M(i',p)-1.
	\end{align} 
	
	Let $k$ and $k'$ be such that $i \in \Bl_k$ and $i' \in \Bl_{k'}$. Let $R$ and $R'$ denote the round-robin orderings constructed in Line $2$ of \Cref{alg:coprime} when called on $(\Bl_k,\Br_m)$ and $(\Bl_{k'},\Br_m)$, respectively. Let $i''$ be the agent whose position in $R$ matches the position of $i'$ in $R'$. Then, $i''$ is matched to the same set of agents from $\Br_m$ during the call to \Cref{alg:coprime} on $(\Bl_k,\Br_m)$ as $i'$ is matched to during the call to \Cref{alg:coprime} on $(\Bl_{k'},\Br_m)$. Because \Cref{alg:coprime} produces an SD-DEF1 matching, \Cref{eqn:desired} holds when $i'$ is replaced by $i''$, and therefore, must also hold for $i'$. 
		
Due to the exact same argument, matching $M$ is also SD-EF1 over $\Nl$, as desired.
\end{proof}	

Finally, we turn our attention to the general case in which we drop the constraints $\nl=\nr$ and $\dl_i=\dr_j=d$. We do however require that $\nl \cdot \dl=\nr \cdot \dr$. The proof of \Cref{thm:general}
 uses a trick of adding dummy agents to the side with fewer agents, computing an SD-DEF1 matching as per \Cref{lem:not-coprime}, and then removing the dummy agents. The key is to show that the removal of dummy agents reduces the degrees of the agents on the opposite side exactly as intended and SD-DEF1 is preserved.
\begin{proof}[Proof of Theorem \Cref{thm:general}]
Without loss of generality, assume that $\nl \le \nr$, and hence, $\dl \ge \dr$. We add $\nr-\nl$ dummy agents with indices $\nl,\ldots,\nr-1$ to the left, so each side has exactly $\nr$ agents. We extend the preferences of the agents on the right so that the dummy agents, indexed higher than the real agents, appear at the bottom of their preference rankings. 

Now, we run the Restricted-Round-Robin algorithm (\Cref{alg:no-coprime}) with inputs $\nr$ and $\dl$. Let $M$ denote the matching returned. Note that while each agent on the left has the intended degree $\dl$, each agent on the right has degree $\dl$, instead of $\dr$. 

Finally, we remove the dummy agents from the left, which reduces the degrees of the agents on the right who were matched to them. Let $\hat{M} = M([\nl],[\nr])$ be the matching $M$ restricted to the real agents. Our goal is to show that $\hat{M}$ is a complete SD-DEF1 matching for the original problem. 

First, note that all agents on the left still have degree $\dl$ under $\hat{M}$. To show that all agents on the right now have degree $\dr$ under $\hat{M}$, we need to show that they are matched to an equal number ($\dl-\dr$) of dummy agents under $M$. Suppose this is not the case. Then, there exist agents $j,j' \in \Nr$ such that $j$ is matched to at least two more dummy agents than $j'$ under $M$. It is easy to check that this violates SD-DEF1 of $M$, which is a contradiction because \Cref{alg:no-coprime} returns an SD-DEF1 matching. Thus, after removing the dummy agents, the degree of all agents on the right drop to precisely $\dr$. Hence, $\hat{M}$ is a complete matching.

To show that $\hat{M}$ is SD-DEF1, note that it is trivially SD-EF1 over the left side because $M$ is SD-EF1 over this side and the matches of these agents do not change. It is also SD-EF1 over the right side because $M$ is SD-EF1 over this side and exactly $\dl-\dr$ least-preferred agents are removed from the allocations of every agent in $\Nr$. 
\end{proof}

We note that it is possible to extend our constructive result slightly beyond the case of $\nl \cdot \dl = \nr \cdot \dr$. Without loss of generality, assume that $\nl \cdot \dl < \nr \cdot \dr$. First, note that in this case, no matching is complete. We can still make the degree of each agent on the left equal to $\dl$, but the best we can hope for is that the degrees of agents on the right differ by at most $1$, i.e., they are either $\floor{\nicefrac{\nl \cdot \dl}{\nr}}$ or $\ceil{\nicefrac{\nl \cdot \dl}{\nr}}$.\footnote{In case that $\nicefrac{\nl \cdot \dl}{\nr}$ is an integer, we can set this to be $\dr$ and achieve exactly equal degrees on the right side too.} In this case, the trick outlined in \Cref{thm:general} only works when the dummy agents are added to the left side, i.e., if $\nl \le \nr$. We conjecture that such an SD-DEF1 matching always exists even when $\nl > \nr$, but leave it as an open question.

%% file: 4-DMMS.tex
In this section, we focus first on the existence of DMMS matchings, and second on the existence of matchings that are DMMS and SD-DEF1 concurrently. 

We begin by considering the case where agents on both sides have identical preferences, i.e., $\ul_i(j) = \ul_{i'}(j)$, for any pair of agents $i,i' \in \Nl$, and any $j \in \Nr$, and similarly  $\ur_j(i) = \ur_{j'}(i)$, for any pair of agents $j,j' \in \Nr$ and any $i \in \Nl$. We show the following negative result, which stands in contrast to the one-sided fair division setting in which an MMS allocation is guaranteed to exist when agents have identical preferences.

\begin{theorem}\label{theor:DMMS-not-guaranteed}
A  $0.89$-DMMS matching may not exist, even when agents on both sides have identical preferences. 
\end{theorem}
\begin{proof}
	We  denote by $\ul$ and $\ur$ the cardinal preferences of the agents in $\Nl$ and $\Nr$ respectively. As the utilities are the same across the agents in the same group, we can define $\mmsl=\mmsl_i$ for all $i \in [n^\ell]$, and $\mmsr=\mmsr_j$ for all $j \in [n^r]$.

	Consider the instance with $n = \nl = \nr =7$ and $d=\dl = \dr =3$, $\ul(j)=n-j-1$ for all $j \in [n]$, and $\ur(i)=n-i-1$ for all $i \in [n]$. Thus, for any complete matching,
	$\sum_{i \in \Nl} \ul (M_i^\ell)=\sum_{j \in \Nr} \ur (M_j^\ell)=63$.
	This means that $\mmsl=\mmsr \leq 9$, because if all agents receive equal utility then they each get utility $9$. Next, we construct a matching $M$ such that $\ul(M^\ell_i)=9$ for all $i\in [n]$.

	Without loss of generality, assume that $M^\ell_0$, $M^\ell_1$, and $M^\ell_2$ all contain agent $0$. Then, we know that agents $1$ and $2$ cannot be contained in these bundles, because then they would have value larger than $9$, implying that some other agent receives utility less than $9$. Without loss of generality, we assume that bundles $M^\ell_3$, $M^\ell_4$, and $M^\ell_5$ contain agent $1$. Now, we observe that  $M^\ell_6=\{2,3,4\}$, as there is no other way to have $\ul(M^\ell_6)=9$. As $0$ and $2$ can not belong to the same bundle (such a bundle would be valued at least 10), we may assume without loss of generality that agent $2$ is contained in $M^\ell_3$, and $M^\ell_4$. Then, the constraint that $\ul(M^\ell_3)=\ul(M^\ell_4)=9$ dictates that $M^\ell_3=M^\ell_4=\{1,2,6\}$.
	With these bundles fixed, it is easy to check that the only $M^\ell_5$ that yields $\ul(M^\ell_5)=9$ is $M^\ell_5=\{1,3,5\}$.
	Lastly, without loss of generality, we may assume that $\Ml_0=\Ml_1=\{0, 4, 5\}$, and $\Ml_2= \{0,3, 6\}$. Hence, we conclude that the following matching is the only one (subject to permutations of $\Nl$) that satisfies MMS for agents on the left.

	    \begin{itemize}
			\item $M^\ell_0=M^\ell_1=\{0, 4, 5\}$
			\item $M^\ell_2=\{0, 3, 6\}$
			\item $M^\ell_3=M^\ell_4=\{1,2,6\}$
			\item $M^\ell_5=\{1, 3, 5\}$
			\item $M^\ell_6=\{2, 3, 4\}$
			\item[\vspace{\fill}]
	    \end{itemize}

	Now, consider agents $0 \in \Nr$ and $4 \in \Nr$. Both are matched to agents $0 \in \Nl$ and $1 \in \Nl$, but agent $0 \in \Nr$ is matched to agent $2 \in \Nl$ while agent $4 \in \Nr$ is matched to agent $6 \in \Nl$. Therefore, $\ur(M^r_0) \neq \ur(M^r_4)$ (and this difference persists regardless of permutations of $\Nl$). It is therefore not the case that every agent on the right receives utility 9; in particular, one agent receives utility 8 or less, producing the approximation ratio $\alpha=8/9<0.89$.
\end{proof}

While a DMMS matching may not exist, even when preferences are identical, we can exploit the algorithms presented in Section~\ref{sec:envy-free} to obtain an approximation to DMMS.

\begin{theorem}\label{theor:SDDEF1-implies-1/dMMS}
	When $\nl \dl = \nr \dr$ and both groups of agents have identical utilities, every complete SD-DEF1 matching $M$ is also $\frac{1}{\dl}$-MMS over $\Nl$, and $\frac{1}{\dr}$-MMS over $\Nr$. 
\end{theorem}
\begin{proof}
Let $M$ be a complete matching that is SD-EF1 over $\Nl$. Let $\ul$ denote the common utility function for agents in $\Nl$ which induces (without loss of generality) the preference ranking $0 \succ \ldots \succ \nr-1$ over $\Nr$. We show that $M$ is also $\nicefrac{1}{\dl}$-MMS over $\Nl$. By symmetry, the MMS approximation for $\Nr$ also follows. 

First, note that each agent in $\Nl$ must be matched to at least one agent in $\set{0,\ldots,\ceil{\nl/\dr}-1} \subset \Nr$. This is because if some agent $i \in \Nl$ is matched to none of these agents, then some other agent $i' \in \Nl$ must be matched to at least two of them, which would violate SD-EF1 over $\Nl$.  

Let $\hat{M}$ be a matching that is MMS over $\Nl$; it is easy to see that under identical cardinal valuations, an MMS matching must exist. Because $(\ceil{\nl/\dr}-2) \cdot \dr \le (\nl/\dr-1) \cdot \dr < \nl$, there must exist $i \in \Nl$ such that $\hat{M}^\ell_i \cap \set{0,\ldots,\ceil{\nl/\dr}-2} = \emptyset$. In other words, if $j^* = \argmax_{j \in \hat{M}^\ell_i} \ul(j)$, then $j^* \ge \ceil{\nl/\dr}-1$. Notice that $\mmsl \le \ul(\hat{M}^\ell_i) \le \dl \cdot \ul(j^*)$ because $|\hat{M}^\ell_i| = \dl$. 

Now, because each agent in $\Nl$ is matched to at least one agent in $\{0,\ldots,\ceil{\nl/\dr}-1\}$, it follows that each agent receives utility at least $\ul(j^*) \geq \nicefrac{1}{\dl}\cdot  \mmsl $, as desired. 
\end{proof}

We next show an almost-matching upper bound that can be achieved by any SD-DEF1 matching, to complement Theorem~\ref{theor:SDDEF1-implies-1/dMMS}. In fact, we show a more general result that trades off the approximation to DMMS with the approximation to double envy-freeness. 

\begin{theorem}\label{theor:incompatibility}
There exists an instance with $n^\ell = n^r = n$ and $d^\ell = d^r = d$ in which no matching is simultaneously $\frac{c+2}{d}$-DMMS and SD-DEF$c$ for any $c \in [d]$.
\end{theorem}
\begin{proof}
We present an example in which there is no  matching that is concurrently $\frac{c+2}{d}$-MMS and SD-EF$c$ over $\Nl$.
Let $n$ be divisible by $d$. We consider the instance in which, for all $i \in \Nl$, $\ul_i(j)=d+1$ for $j \in \{0,...,n/d-2\}$,  $\ul_i(j)=1$ for $j \in \{n/d-1,...,n/d+d-2\}$, and $\ul_i(j)=0$ for $j \in \{n/d+d-1,...,n\}$. Hence, $\mmsl$ is equal to $d$, as $n-d$ agents of $\Nl$ can be matched  with the first $n/d-1$ agents of $\Nr$ and take utility equal to $d+1$, and the remaining $d$ agents can have $d$ matches in the set $\{n/d-1,...,n/d+d-2\}\subset \Nr$ and receive utility equal to $d$. 
Because there are less than $n+d^2$ matches available in the subset $\{0,...,n/d+d-2\} \subset \Nr$, and $n \geq d^2$, it must be the case that there are agents on the left that have at most one match to this subset of $\Nr$. 
So, to achieve a SD-EF$c$ allocation with respect to the agents in $\Nl$, each agent can have at most $c+1$ matches to this subset. But then only $n-d$ out of $n$ agents have utility at least $d+1$, and the remaining ones have utility at most $c+1$. So, there are agents that receive utility at most $c+1$ while $\mmsl=d$.
\end{proof}	

In \Cref{app:theor:def1-and-dmms}, we establish the existence of a complete SD-DEF1 and DMMS matching by restricting the degree bound.

Finally, we show that a strong impossibility persists even if we only require SD-EF1 on one side and MMS on the other.

\begin{theorem}\label{theor:EF1-MMS-impossibility}
	A matching that satisfies SD-EF1 over $\Nl$ and MMS over $\Nr$ is not guaranteed to exist, even when agents on both sides have identical preferences.
\end{theorem}
\begin{proof}
We consider the instance in which $n^\ell=n^r=n=11$, and $d^\ell=d^r=d=3$, while $u^\ell=u^\ell_i(j)=n-j-1$ for all $i \in N^\ell$ and $j \in N^r$ and $u^r=u^r_j(i)=n-i-1$ for all $i \in N^\ell$  and $j \in N^r$. Finding  $MMS^\ell$ can be done using the following integer linear
program $ILP_1(n,d, v^\ell)$:
\begin{equation*}
\begin{array}{ll@{}r}
\text{maximize}  &  MMS^\ell &\\
\vspace*{0.2cm}
\text{subject to}&  \sum_{j \in  [n]}  M(i,j) \cdot v^\ell(j) \geq MMS^\ell,  & \forall i \in [n]\\
\vspace*{0.2cm}
& \sum_{j \in [n]}  M(i,j)  = d,  & \forall i \in [n]\\
\vspace*{0.2cm}
&  \sum_{i \in  [n]}  M(i,j)  =d,  & \forall j \in[n]\\
\vspace*{0.2cm}
  &      M(i,j) \in \{0,1\}, & \forall i,j \in [n]
\end{array}
\end{equation*}
The first constraint ensures that the matchings is MMS over $N^\ell$, and the remaining ones ensure that the matching is valid and complete. 

Detecting whether it exists a matching that is SD-EF1 over $N^r$ and  MMS over $N^\ell$ can be done using the following integer linear program $ILP_2(n,d, v^\ell, MMS^\ell)$:
\begin{equation*}
\begin{array}{l@{}lr}
 \vspace*{0.3cm}
\sum_{i \in  [r]}  M(i,j)  \geq \sum_{i \in  [r]}  M(i,j')-1, & & \forall j, j' \in [n], \forall r \in [n]\\
 \vspace*{0.3cm}
\sum_{j \in  [n]}  M(i,j) \cdot v^\ell(j) \geq MMS^\ell, &  & \forall i \in [n]\\
\vspace*{0.3cm}
\sum_{j \in [n]}  M(i,j)  = d, & & \forall i \in [n]\\
\vspace*{0.3cm}
 \sum_{i \in  [n]}  M(i,j)  =d, & & \forall j \in[n]\\
 \vspace*{0.3cm}
M(i,j) \in \{0,1\},& & \forall i,j \in [n]
\end{array}
\end{equation*}
The first constraint ensures that the matchings is SD-EF1 over $N^r$, the second one that it is MMS over $N^\ell$, and the remaining ones ensure that the matching is valid and complete.

Using the above ILP programs, we check that the aforementioned instance does not admit a matching that is concurrently SD-EF1 over $N^r$ and MMS over $N^\ell$. 

\end{proof}

%% file: 6-discussion.tex
We have introduced a model that bridges two-sided matching and fair division by requiring fairness on both sides of a matching market. We have shown that SD-EF1 can be achieved for agents on both sides, when all agents on the left side  and all agents on the right share a common ordinal preference ranking over agents on the other side. When this condition is not satisfied, there may exist no matching that satisfies SD-DEF1. We have also shown that there may not exist a doubly MMS matching even when agents have identical preferences. While we do not rule out a good approximation to DMMS, we show that it is essentially impossible to obtain a good approximation to DMMS if one also requires SD-DEF1.

It is interesting to note that the proofs of Theorems \ref{theor:SDDEF1-implies-1/dMMS} and \ref{theor:incompatibility} do not rely on the constraints that an agent in $\Nl$ can have up to $d$ matchings, and can be matched with an agent in $\Nr$ at most once. Therefore, these theorems also hold in a version of the one-sided fair division problem where there are $n$ agents and $n/d$ items with $d$ copies each, and all the agents have identical preferences. 

Many interesting avenues for future research remain. For example, one can hope to derive weaker positive results in the case where one side has identical preferences. When the side with identical preferences does the picking, \Cref{alg:coprime} remains EF1 for that side. In \Cref{app:simulations}, we conduct empirical simulations and observe that \Cref{alg:coprime} remains EF1 for some of the agents on the other side as well (and does better than the classical round-robin algorithm in this aspect). It would also be interesting to compare the two-sided fair division setting with its one-sided counterpart (where only one side has preferences and we seek fairness only for this side). In our simulations, we observe that there is a sharp contrast for envy-freeness (one-sided EF is almost always achievable while two-sided DEF almost always isn't). For the maximin share guarantee, however, there is no contrast: both one-sided MMS and two-sided DMMS are almost always achievable. Theoretically analyzing the probability of satisfiability of these notions in random instances would be an interesting direction for the future. One can also consider two-sided versions of other fairness notions, including those that remain interesting when the degree constraint is 1, which could yield further interesting results in the one-to-one or many-to-one settings. Finally, it would also be interesting to derive positive results when each agent can have a different degree constraint. 

%% file: app-1-arb_pref_d=2.tex
We show the following theorem.

\begin{theorem}\label{theor:arb_pref_d=2}
When $\nl=\nr=n$, $\dl=\dr=2$, and at least one group of agents has identical ordinal preferences, a complete SD-DEF1 matching always exists and can be computed in polynomial time. 
\end{theorem}

The proof follows from two lemmas; one of which treats the case where $n$ is even, and one where $n$ is odd.

\begin{algorithm} [htb!]
	\caption{Three-Phase-Round-Robin-I ($n,  d, \pi^r $ ) }\label{alg:arb_prefer_even}
	\begin{algorithmic}[1]
		\vspace*{2mm}
		\Statex \textit{Phase 1:}\dotfill
		\vspace*{1mm}
		\For { $j=0,\ldots,\nicefrac{n}{2}-1 $}
		\State Match agent $j$ on the right to her most preferred agent on the left with no existing matches.
		\EndFor
		\vspace*{2mm}
		\Statex \textit{Phase 2:}\dotfill
		\vspace*{1mm}
		\For{$j= \nicefrac{n}{2},\ldots,n-1$ }
		\State Match agent $j$ on the right with her most preferred agent on the left which already has one match. 
		\EndFor
		\vspace*{2mm}
		\Statex \textit{Phase 3:}\dotfill
		\vspace*{1mm}
		\For { $j= n-1,\ldots,\nicefrac{n}{2}$}
		\State Let $i'$ be the agent on the left that agent $j$ on the right is already matched to.
		\State Let $j' \neq j$ be the other agent on the right that $i'$ is matched to ($j'$ must exist).
		\State Match both $j$ and $j'$ to agent $j$'s most preferred agent on the left who has no existing match.
		\EndFor
	\end{algorithmic}
\end{algorithm}

\begin{lemma}\label{lemma:arb_pref_d=2_even}
When $d=2$ and $n$ is even, Algorithm \ref{alg:arb_prefer_even} returns a complete  SD-DEF1 matching.
\end{lemma}
\begin{proof}	
	 
First, observe that any agent in $\Br_1$, and any agent in $\Br_2$ have one match in $\Bl$ and one match in $\Nl\setminus \Bl$. Denote with $y_j$  and $z_j$ the matches of agent $j \in \Nr$ in $\Bl$ and  in $\Nl\setminus \Bl$, respectively. Moreover notice that for every agent $j \in \Br_1$ there is an agent $j' \in \Br_2$, such that $y_j=y_{j'}$ and $z_j=z_{j'}$, as in Phase $3$ when $j'$ is matched to $z_j$, the same agent is also matched to $j$ if $y_j=y_{j'}$.
 
 We can easily verify that the matching is SD-EF1 over  $\Nl$, as each agent in $\Nl$ has one match in $\Br_1$, and one match in $\Br_2$. Now, we prove that the matching is also SD-EF1 with respect to the agents in $\Nr$.  We consider the three following cases.

\medskip\noindent\textbf{Case 1: $j,j' \in \Br_1$.}
 Without loss of generality, we assume that $j<j'$. Then,  $y_j \succ_{\rr_j} y_{j'}$, as $j$ chooses before $j'$ in Phase $1$. Thus, $j$ cannot envy $j'$ for more than one matches. Moreover, as $j$ has one match in $\Bl$, and one match $\Nl\setminus \Bl$, and as $j'$ prefers $z_{j'}$ to any agent in $\Nl\setminus \Bl$ (otherwise she could have chosen an agent in $\Nl\setminus \Bl$ as in Phase $1$ all of them have no matches), we conclude that $j'$ does envy $j$ for more that one matches.

\medskip\noindent\textbf{Case 2: $j,j' \in \Br_2$.}
 Without loss of generality, we assume that $j<j'$. Then, $y_j \succ_{\rr_j} y_{j'}$, as $j$ appears before $j'$ in Phase 2. Thus, $j$ cannot envy $j'$ for more than one matches. On the other hand, $z_{j'} \succ_{\rr_{j'}} z_{j}$, as $j'$ appears before $j$ on Phase 3. Hence $j'$ does envy $j$ for more that one matches.

\medskip\noindent\textbf{Case 3: $j \in \Br_1$ and  $j' \in \Br_2$.}
	If $j$ and $j'$ share the same matches, then obviously they don't envy  each other. Otherwise, we denote with $\hat{j}$ the agent in $\Br_2$  that has the same matches with $j$, and with $\hat{j}'$ the agent in $\Br_1$ that has the same matches with $j'$. Then, the theorem follows from cases $2$ and $3$, as the matching is SD-EF1 with respect to $j$ and $\hat{j}$, and with respect to $j'$ and $\hat{j}'$.
\end{proof}

\begin{algorithm} [h]
	\caption{Three-Phase-Round-Robin-II ($n, d, \pi^r $) }\label{alg:arb_prefer_odd}
	\begin{algorithmic}[1]
		\vspace*{2mm}
		\Statex \textit{Phase 1:}\dotfill
		\vspace*{1mm}
		\For {$j=0,\ldots,\ceil{n/2}-1$}
		\State Match agent $j$ on the right to her most preferred agent on the left with no existing matches.
		\EndFor
		\vspace*{2mm}
		\Statex \textit{Phase 2:}\dotfill
		\vspace*{1mm}
		\For{$j=\ceil{n/2},\ldots,n-2$}
			\State Match agent $j$ on the right with her most preferred agent on the left which already has one match. 
		\EndFor
		\State Let $i'$ be the  agent on the left that agent $\ceil{n/2}-1$ on the right is matched to.
		\If{$i'$ has only one match}
				\State Match agent $n-1$ on the right to agent $i'$ on the left.
		\Else
				\State Match agent $n-1$ on the right to her most preferred agent on the left with no existing matches.
		\EndIf		
		\State Match agent $\ceil{n/2}-1$ on the right with agent $i'$ on the left with exactly one existing match (such an $i'$ must exist)
		\vspace*{2mm}
		\Statex \textit{Phase 3:}\dotfill
		\vspace*{1mm}
		\For {$j= n-1,\ldots,\ceil{n/2}$}
			\State Let $i'$ be the agent on the left that agent $j$ on the right is already matched to.
			\State Let $j' \neq j$ be the other agent on the right whose only match is to agent $i'$.
			\State Match both agents $j$ and $j'$ to agent $j$'s most preferred agent on the left who has no existing matches.
		\EndFor
		\State Match the agent on the right with one existing match, to the agent on the left with one existing match.
	\end{algorithmic}
\end{algorithm}

\begin{lemma}\label{lemma:arb_pref_d=2_odd}
When $d=2$ and $n$ is odd, Algorithm \ref{alg:arb_prefer_odd} returns a complete  DEF1 allocation.
\end{lemma}
\begin{proof}
 Intuitively, the algorithm works in three Phases. Let us divide $\Nr$ into two sets: $\Br_1=\{0,...,ceil{n/2}-1\}$, and  $\Br_2=\{\ceil{n/2},...,n-1 \}$. In the first Phase, one-by-one we match agents in $\Br_1$ to their most preferred agent on the left who has no prior matches. We do so in the best-to-worst order over agents in $\Br_1$ according to $\rl$. Let $\Bl$ be the set of agents in $\Nl$ who now have degree $1$; note that $|\Bl| = \ceil{n/2}$. 

In the second Phase, we repeat a similar process, except with agents in $\Br_2\setminus n-1$, still in the best-to-worst order according to $\rl$, and by matching them to their most preferred agent in $\Bl$. Then if $i'$ on the left side, with whom $\ceil{n/2}-1$ on the right is matched to, has only one match,  $n-1$ on the right side is matched with her, otherwise $n-1$ is matched with her best choice in $\Bl$. At this point,  $\floor{n/2}$ agents in  $\Bl$ have degree $2$, and only has degree $1$, and hence $\ceil{n/2}-1$ agent on the right is matched to agent in $\Bl$ with degree $1$. So,  at the end of this Phase, all agents on the right have degree $1$, except for $\ceil{n/2}-1$ that has degree $2$, while agents in $\Bl$ have degree $2$ and agents in $\Nl\setminus\Bl$ have degree $0$. 

In the last Phase, we again consider agents in $\Br_2$, but now in the \emph{worst-to-best} order according to $\rl$. Note that each such agent $j$ is already matched to some agent $i' \in \Bl$, and agent $i'$ is also matched to some agent $j'$ from $\Br_1$. Note that all such $j'$ from $\Br_1\setminus \ceil{n/2}-1$ has degree $1$. Hence, if $j' \neq \ceil{n/2}-1$ both $j$ and $j'$ to the most preferred agent of $j$ from $\Nl\setminus\Bl$, otherwise only $j$ is matched to her. At the end, there is one $j$ in $\Br_1$ , and one agent $i$ in $\Nl\setminus \Bl$ that have degree one, and we matched them.

First, observe that any agent in $\Br_1\setminus \ceil{n/2}-1$, and any agent in $\Br_2$ has one match in $\Bl$ and one match in $\Nl\setminus \Bl$,  while $\ceil{n/2}-1$ has two matches in $\Bl$.  Denote with $y_j$  and $z_j$ the matches of agent $j \in \Nr\setminus \ceil{n/2}-1$ in $\Bl$ and  in $\Nl\setminus \Bl$, respectively.

We denote with $j^*$ the agent in $\Br_2$ that share the same match on the left with $\ceil{n/2}-1$ on the right and with $\hat{j}^*$ the agent in $\Br_1$ that share the same match on the left with $j^*$. We see that for every agent $j' \in \Br_2\setminus j^*$ there is an agent $j \in \Br_2$, such that $y_j=y_{j'}$ and $z_j=z_{j'}$, as in the third Phase when $j'$ is matched to $z_j$, the same agent is also matched to $j$ if $y_j=y_{j'}$.
 
 We can easily verify that the matching is SD-EF1 over  $\Nl$, as each agent in $\Nl$ either has one match in $\Br_1$, and one match in $\Br_2$, or has two matches  in $\Br_1$ but one of them is with $\ceil{n/2}-1$.
 
 Now, we prove that the allocation is also SD-EF1 with respect to the agents in $\Nr$.  We consider the three following cases.

\medskip\noindent\textbf{Case 1: $j,j' \in \Br_1$.}
This case is similar as Case 1 of the proof of lemma \ref{lemma:arb_pref_d=2_even}, so we omit the details.
 
\medskip\noindent\textbf{Case 2: $j,j' \in \Br_2$.}
This case is similar as Case 2 of the proof of lemma \ref{lemma:arb_pref_d=2_even}, so we omit the details.

\medskip\noindent\textbf{Case 3: $j \in \Br_1$ and  $j' \in \Br_2$.}
 First we know $j$ prefers $y_j$ to any agent in $\Nl\setminus \Bl$, and as $j'$ has only one match in $\Bl$, we conclude that $j$ does not envy $j'$ for more than one matches. 

We denote and with $\hat{j}$ the agent in $\Br_1$ that has the same matches with $j' \in \Br_2 \setminus j^*$.

We assume that $j\neq \{\ceil{n/2}-1,\hat{j^*}\}$. Then, we know from case $2$ that $j'$ does not envy  $\hat{j}$ for more than one match, and similar $j$.  If $j= \ceil{n/2}$, then either $j'$ and $j$ share the same matches in $\Bl$, or $j'$ before line 10 has chosen a better agent rather than the one that $j$ is matched to in this line.

Next, if  $j =\hat{j}^*$ and  $j' = j^*$, $j$ and $j'$ share one match, and  $j'$ does not envy $j$ for more than one match, while if $j' \neq j^*$ we know that  $j'$ prefers $y_{j'}$ to $y_{\hat{j}^*}$, as  $y_{\hat{j}^*}$ has only one match before line 10. 
\end{proof}

\begin{proof}[Proof of \Cref{theor:arb_pref_d=2}]
	The proof follows from \Cref{lemma:arb_pref_d=2_even,lemma:arb_pref_d=2_odd}.	
\end{proof}

%% file: app-4-SD-DEF1-MMS-d=2.tex
For $d \ge 3$, Theorem~\ref{theor:incompatibility} rules out the possibility of a  matching that satisfies SD-DEF1 and DMMS. In the next theorem, we show that the two properties can be achieved simultaneously when $d=2$. 

\begin{theorem}\label{theor:def1-and-dmms}
	For $d=2$, a matching satisfying SD-DEF1 and DMMS exists and can be computed efficiently, when agents on both sides have identical preferences. 
\end{theorem}
\begin{proof}
	Consider the following matching $M$.  When $n$ is even $\Ml_i=\{i,n-i-1\}$ for $i \in [d]$, and when $n$ is odd $\Ml_i=\{i,n-i-1\}$ for $i\in [d]\setminus \{\floor{n/d}-1,\floor{n/d},\ceil{n/d}\}$, while $\Ml_{\floor{n/d}-1}=\{\floor{n/d}-1, \floor{n/d}\}$, $\Ml_{\floor{n/d}}=\{\floor{n/d}-1, \ceil{n/d}\}$ and $\Ml_{\ceil{n/d}}=\{\floor{n/d}, \ceil{n/d}\}$. Then, $\Ml_i=\Mr_j$, when $i=j$. It is easy to verify that this matching is SD-DEF1. 
	
Let $\Ml_{min}=\argmin_{\Ml_i \in \Ml} \ul(\Ml_i)$. We show that $\ul(\Ml_{min})= \mmsl$. 	Assume for contradiction that there is another  $\hat{M}^{\ell}$, such that $\ul(\hat{M}^{\ell}_{min} )> \ul(\Ml_{min})$, where $\hat{M}^{\ell}_{min}=\argmin_{\hat{M}^{\ell}_i \in \hat{M}^{\ell}} \ul(\hat{M}^{\ell}_i)$. Consider the cases that  $\Ml_{min}=\{i,n-i+1\}$, with $i\leq  \floor{n/d}-2$ when $n$ is odd and $\Ml_{min}=\{i,n-i+1\}$, with $i\leq  n/d-1$, when $n$ is even.	In order to have $\ul(\hat{M}^{\ell}_{min} )> \ul(\Ml_{min})$, all the agents that are at least equal to $n-i+1$ should be included  in a  $\hat{M}^{\ell}_i$ with an agent smaller than $i$, as otherwise the minimum value is not increased. But then  notice that there are $2 \cdot i$ matches in one side that should be matched with $2(i-1)$ matches of the other side, which is impossible. Next, if $\Ml_{min}=\{\floor{n/d}-1, \floor{n/d}\}$, or $\Ml_{min}=\{\floor{n/d}-1, \ceil{n/d}\}$, then all the agents that are at least equal to $\ceil{n/d}$ should be included  in a  $\hat{M}^{\ell}_i$ with an agent smaller than $\floor{n/d}-1$ which is impossible. Lastly, if $\Ml_{min}=\{\floor{n/d}, \ceil{n/d}\}$, then all  the agents that are at least equal to $\floor{n/d}$ should be included  in a  $\hat{M}^{\ell}_i$ with an agent at most equal to  $\floor{n/d}-1$ which is again impossible. Hence, $\ul(\Ml_{min})= \mmsl$.

With similar arguments it can be proved that $\ur(\Mr_{min})= \mmsr$, where $\Mr_{min}=\argmin_{\Mr_j \in \Mr} \ul(\Ar_j)$.	
\end{proof}

%% file: 5-simulations.tex
\begin{figure*}[t] 
\centering
\includegraphics[width=\textwidth]{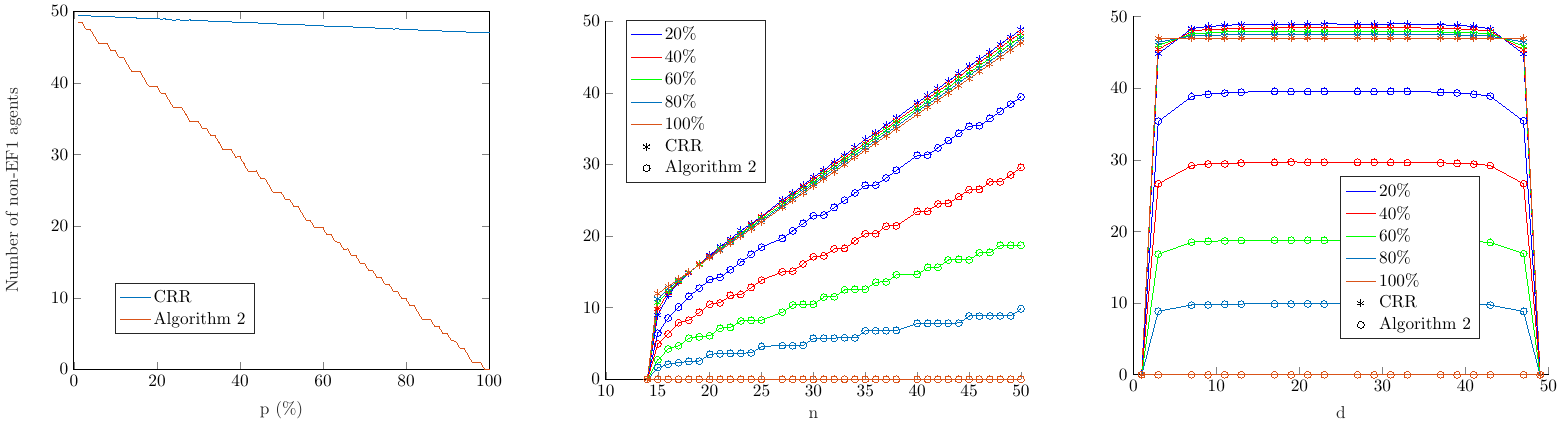}
\caption{Non-EF1 agents for (a) $n=50$, $d=23$ and various values of $p$ (left), (b) $d=13$ and various values of $n$ and $p$ (middle), and (c) $n=50$ various values of $d$ and $p$ (right). }
\label{fig-CRR-DRR}
\end{figure*}

\Cref{sec:possibility} shows that \Cref{alg:coprime} finds an SD-DEF1 matching when \emph{both sides} have identical ordinal preferences. In this section, we conduct experiments to see how well the algorithm works beyond this restricted case. Specifically, we simulate instances in which we start from identical preferences on both sides but, for some agents in $\Nl$ and/or some agents in $\Nr$, replace their preferences by those drawn from the uniform distribution (impartial culture). We measure the number of agents for whom the resulting matching is not EF1 and use the classic round robin (CRR) algorithm as benchmark. 

First, we observe that when even a few agents in $\Nl$ --- the side that picks --- have heterogeneous preferences, both \Cref{alg:coprime} and CRR have almost similar performance. This is quite intuitive as the proof of \Cref{theor:coprime} relies on a delicate mathematical formula for the set of agents picked by each agent in $\Nl$, and even a single agent in $\Nl$ not picking as expected breaks this.

Next, we restrict our attention to instances where the agents in $\Nl$ and $p\%$ of agents in $\Nr$ have identical preferences. In this case, \Cref{alg:coprime} is guaranteed to satisfy SD-EF1 for all agents except the $(100-p)\%$ agents in $\Nr$ with their preferences drawn from the uniform distribution. \Cref{fig-CRR-DRR} shows the number of agents for whom EF1 is violated for different values of $p$ when (a) $n=50$ and $d=27$ (left), (b) $n \in \set{14,\ldots,50} $ and $d=13$ (middle), and (c) $d \in \set{1,\ldots,49}$ and $n=50$ (right).\footnote{We restrict attention to coprime $(n,d)$, so some entries in the plots are skipped.} Each point is an aggregation over $1000$ random instances.

First, we observe that when $p$ is close to $0\%$, CRR and \Cref{alg:coprime} have similar performance, but as $p$ increases, the performance of \Cref{alg:coprime} improves substantially (achieving SD-DEF1 when $p=100\%$) while that of CRR improves negligibly. Quite interestingly, the number of non-EF1 agents does not change significantly with $d$ (except for extremely small or large values of $d$). 

We also conducted simulations to compare the two-sided fair division setting in which both sides have preferences and we seek EF/approximate MMS for both sides to the one-sided fair division setting in which only one side has preferences and we want EF/approximate MMS only for that side. We vary $n \in \set{3,\ldots,8}$ and $d \in \set{1,\ldots,n-1}$, and, for each combination, generate $500$ random instances with additive utilities where the utility of each agent for each good/agent is selected uniformly at random from $\set{0,\ldots,20}$. 

For envy, we find a sharp contrast between one-sided EF and two-sided DEF for $n \ge 6$ and $2 \le d \le n-2$: while EF is almost always achievable (in at least $92.6\%$ of instances) while DEF is almost always unachievable (achievable in at most $7.6\%$ instances).

However, for maximin share guarantee, we do not find such a contrast. One-sided (exact) MMS was always achievable in all our simulations. Two-sided (exact) DMMS was also achievable in at least $92.6\%$ instances, and in the remaining, $0.99$-DMMS was still achievable. 
 